\documentstyle[12pt,epsf]{article}
\begin{document}

\begin{titlepage}
\samepage{
\setcounter{page}{1}

\rightline{UFIFT--HEP--97--3}
\rightline{ANL-HEP-PR-97-11}
\vfill
\begin{center}
 {\Large \bf  Spin Dependent Drell-Yan beyond Leading Order: 
Non-Singlet corrections to $O(\alpha_s^2) $ \\}

\vfill
\vfill
 {\large Sanghyeon Chang$^{*}$\footnote{
        E-mail address: schang@phys.ufl.edu},
        Claudio Corian\`{o}$^{*}$\footnote{
        E-mail address: coriano@phys.ufl.edu},
               R. D. Field$^{*}$\footnote{
        E-mail address: rfield@phys.ufl.edu} \\ and
        L. E. Gordon$^{\dagger}$\footnote{E-mail address: gordon@hep.anl.gov}}
\\
\vspace{.12in}
 {\it $^{*}$   Institute for Fundamental Theory, Department of Physics, \\
        University of Florida, Gainesville, FL 32611, 
        USA\\}
\vspace{.12in}
 {\it $^{\dagger}$  Argonne National Laboratory\\
       9700 South Cass, IL 60439, USA}

\end{center}
\vfill
\begin{abstract}
We present parton-level analytical results for 
the next-to-leading order non-singlet virtual and real corrections to the 
Drell-Yan differential cross-section.  
The dependence of the differential cross section on the helicity of the initial state partons is shown explicitly (the spins of the final state partons are summed). The calculation is 
implemented in dimensional regularization within the $\overline{MS}$ scheme 
and with the t'Hooft Veltman prescriptions for the 
n-dimensional $\gamma_5$. Both the polarized initial state and the unpolarized cross sections can be obtained from our result. 
Our unpolarized cross section agrees with the previous result of
Ellis, Martinelli and Petronzio in the non-singlet sector.

\end{abstract}
\smallskip}
\end{titlepage}

\setcounter{footnote}{0}

\def\beq{\begin{equation}}
\def\eeq{\end{equation}}
\def\beqn{\begin{eqnarray}}
\def\eeqn{\end{eqnarray}}

\def\ie{{\it i.e.}}
\def\eg{{\it e.g.}}
\def\half{{\textstyle{1\over 2}}}
\def\third{{\textstyle {1\over3}}}
\def\quarter{{\textstyle {1\over4}}}
\def\m{{\tt -}}

\def\p{{\tt +}}

\def\slash#1{#1\hskip-6pt/\hskip6pt}
\def\slk{\slash{k}}
\def\GeV{\,{\rm GeV}}
\def\TeV{\,{\rm TeV}}
\def\y{\,{\rm y}}

\def\l{\langle}
\def\r{\rangle}

\setcounter{footnote}{0}
\newcommand{\beqa}{\begin{eqnarray}}
\newcommand{\eeqa}{\end{eqnarray}}
\newcommand{\eps}{\epsilon}

\section{Introduction}
Spin physics has emerged in the last few years as one of the most 
active fields of perturbative QCD. Given the planned experiments at RHIC and HERA, it is likely that such interest will continue to grow. 
Among the interesting processes which are planned to be investigated at 
hadron colliders such as RHIC are the Drell-Yan production of muon pairs
and two-jet production in polarized proton-proton collisions \cite{ramsey}.
The difficulty of calculating spin dependence in next-to-leading order QCD is notorious and therefore only $O(\alpha_s)$ results have been presented in the literature. 

The difficulty in the calculations comes from the presence of chiral fermions or gluons of fixed helicity in the initial state which renders the study of the spin dependencies extremely time consuming. 
The analysis of the infrared structure of the cross section in perturbation 
theory is also much more complex than the unpolarized case. If dimensional regularization is used to control both the infrared and ultraviolet singularities of the amplitudes, then the presence of chiral projectors for 
the fermions requires a suitable definition of $\gamma_5$ in 
n-dimension. 

In this work we present results for the $O(\alpha_s^2)$ virtual corrections to the non-singlet sector of the classical Drell-Yan process.  This is an important ingredient to the overall non-singlet differential cross section, $d
\sigma/d Q^2 d\,td\,u$, and for the total cross section $\sigma(Q^2)$,  
where $Q^2$ is the invariant mass of the photon.  The differential cross section is used to predict the transverse momentum distribution of the 
photon in the process $p\,\, p \to \gamma^* + X $ with polarized beams. 

It has been observed \cite{Jaffe1}
that Drell-Yan production of charged lepton pairs is a case where transverse and longitudinal asymmetries are comparable. To leading order, 
the ratio between the transverse and the longitudinal asymmetries 
$R\equiv A_{TT}/A_{LL}$ is estimated to be of order 1. 
This result is due to the fact that the gluons don't couple to transverse asymmetries and to order $\alpha_s$ only the quark-antiquark channel contribute. Furthermore, Drell-Yan is one of the possible avenues to obtain information about the 
twist-2 transversity distribution, which can be related to the
ratio between transverse and longitudinal asymmetries. 
The inclusion of polarization in the initial states can be a useful tool 
to investigate such effects, and other effects, such as 
compositeness and supersymmetry (see \cite{bour} and references therein). 
The presence of a massive photon in the final state makes an analytic 
calculation much more involved than, for instance, two other recent 
similar studies \cite{CG}. Although the non-singlet contribution to the process can only give information about the polarized quark distributions, it is the most lengthy Drell-Yan subprocess. We present here a brief account of 
the result for the order $\alpha_s^2$ corrections to the non singlet sector.

\section{General Structure }

The lowest order ({\it non-singlet}) contributions to the large 
transverse momentum production of virtual photons  (see Fig.~1) with 
invariant mass $q^2=Q^2$ arise from the two, $q+\bar{q}\to 
\gamma^*+g$,  ``Born" amplitudes shown in Fig.~3.  We refer 
to these two diagrams as the {\it direct} and the {\it exchange} 
(or crossed) amplitudes, respectively. 
In Fig.~2 we have generically illustrated the expansion of the 
amplitude which appears in $d\sigma/d\,Q^2$ up to order $\alpha_s$. Notice that 
the quark form factor contributions represented 
in Fig.~3 (and the related $O(\alpha_s^2)$ 
corrections, not included in the picture) 
do not appear in the study of the cross section, 
$d\sigma/d^2 q_t d\,y$, for transverse momentum, $q_t$, greater than zero. 
The two-loop $O(\alpha_s^2)$ corrections 
to the quark form factor can be added in the study of 
the total cross section, $d\sigma/d\,Q^2$, by interfering the 2-loop on-shell quark form factor 
of ref. \cite{formfac} with the lowest order $q\bar{q}\gamma$ annihilation channel and by using helicity projectors for the initial quark states.

We calculate the spin dependence of 
the cross section by using the helicity projectors, 
$P_{\pm}={1\over 2}(1 \pm \gamma_5)$, which project out 
the helicity states of an initial state quark and antiquark, 
respectively, as follows:
\beq
u(p_1,h_1)={1\over 2}(1+h_1\gamma_5)u(p),\qquad
\bar{v}(p_2,h_2)={1\over 2}\bar{v}(p_2)(1+h_2 
\gamma_5),
\eeq
where $h_1=\pm 1=\pm 2 \lambda_1$ corresponds to quark helicity 
$\pm {1\over 2}$, and $h_2=\pm 1=\pm 2 \lambda_2$ corresponds to antiquark 
helicity $\pm {1\over 2}$.

The squares of the direct amplitude $L_1$ and exchange amplitude 
$L_2$ in Fig.~3 in $N=4-2\epsilon $ dimensions are given by
\beqn
&& M_{dd}(h_1,h_2)=e_f^2g^2g_s^2
{C_F\over N_c}
{2u\over t}\left( (1-\eps)^2 -h_1h_2(1+\eps)^2\right),\nonumber \\
&& M_{cc}(h_1,h_2)=e_f^2g^2g_s^2
{C_F\over N_c}
{2t\over u}\left( (1-\eps)^2 -h_1 h_2(1+\eps)^2\right).
\eeqn
where $\alpha=g^2/4\pi$ is the fine structure constant, and 
$e_f$ the charge of the quark.  The quantity $C_F/N_c$ is the 
color factor, and $\alpha_s=g_s^2/4\pi$ is the QCD strong 
coupling constant. The interference term is more complicated 
and is given by,
\beqn
&& 2M_{dc}(h_1,h_2)= \nonumber \\
&& e_f^2g^2g_s^2
{C_F\over N_c}
{4\over tu}\left[ (1-\eps)(Q^2s-\eps tu)\right.
\left.-h_1h_2(1+\eps)(Q^2s+\eps tu) -2h_1h_2\eps 
tu\right].
\eeqn
The sum of the two Born amplitudes squared is 
\beqn
&& |M_{B}(h_1,h_2)|^2=|M_{B}(h)|^2=M_{dd}(h)+2M_{dc}(h
)+M_{cc}(h)\nonumber \\
&& =e_f^2g^2g_s^2
{C_F\over N_c}
{2\over tu}\left[(1-\eps)\left(2Q^2s+(1-\eps)(t^2+u^2)-2\eps 
tu\right)\right.\nonumber \\
&&\left.-h(1+\eps)\left(2Q^2s+(1+\eps)(t^2+u^2)+2\eps tu\right) 
-4h\eps tu\right],
\eeqn
and depends only on the product $h=h_1h_2$.  

To any order in perturbation theory we can write,
\beq
|M(h_1,h_2)|^2=|M(h)|^2=(1-h)|\bar M|^2 +h|M_{++}|^2 
=|\bar M|^2 +h\Delta |M|^2,
\eeq
where $h=h_1h_2=4\lambda_1\lambda_2$ and where
\beq
|\bar M|^2={1\over4} 
|\sum_{h_1,h_2}M(h_1,h_2)|^2={1\over4}
\sum_{h_1,h_2}|M(h_1,h_2)|^2,
\eeq
is the spin averaged ({\it unpolarized}) amplitude squared.  
Furthermore,
\beq
|M_{--}|^2=|M_{++}|^2\ {\rm and}\  |M_{-+}|^2=|M_{+-}|^2,
\eeq
so that
\beq
|\bar M|^2={1\over2} \left( |M_{++}|^2+|M_{+-}|^2\right).
\eeq
The spin asymmetry, $\Delta|M|^2$, is defined according to
\beq
|M_\Delta|^2={1\over2}\left( |M_{++}|^2-
|M_{+-}|^2\right)
=|M_{++}|^2-|\bar M|^2.
\eeq

The spin averaged ({\it unpolarized}) amplitude squared is 
determined from $|M(h)|^2$ by setting $h=0$ and 
$\Delta|M|^2$ is the coefficient of $h$.  For the Born term, this results 
in
\beq
|\bar M|^2=e_f^2g^2g_s^2
{C_F\over N_c}
{2\over tu}\left[(1-\eps)\left(2Q^2s
+(1-\eps)(t^2+u^2)-2\eps tu\right)\right],
\eeq
and
\beq
\Delta|M|^2=-e_f^2g^2g_s^2
{C_F\over N_c}
{2\over tu}\left[(1+\eps)\left(2Q^2s +(1+\eps)(t^2+u^2)+2\eps 
tu\right)+4\eps tu\right].
\eeq
Adding these two terms yields
\beq
|M_{++}|^2=-e_f^2g^2g_s^2
{C_F\over N_c}
{8\eps\over tu}\left[Q^2s+(t+u)^2\right],
\eeq
which is proportional to $\eps$ and vanishes in the limit 
$\eps\rightarrow 0$.  Since at the Born level there are no 
${1\over\eps}$ singularities that might combine with this term 
to yield a finite contribution, in the limit $\eps\rightarrow 0$,
\beq
|M_{++}|^2=0\ {\rm and}\ \Delta |M|^2=-|\bar M|^2.
\label{spin1}
\eeq
For the $q+\bar{q}\to \gamma^*+g$ subprocess the condition that
$|M_{++}|^2=0$ means that the quark helicity is 
maintained (does not flip) in the collision. The incoming
quark line with helicity $\pm {1\over 2}$ turns around and becomes an
incoming antiquark with helicity $\mp {1\over 2}$.  We refer to this as ``helicity conservation".  We see that at the Born level helicity is
conserved in the limit $\eps\rightarrow 0$.

In $N=4-2\eps$ dimensions  the differential cross section is 
related to the $2\to 2$ invariant amplitude according to
\beq
s{d\sigma\over dt}(s,t,h)={1\over 16\pi s}
\left({4\pi s\over tu}\right)^\eps{1\over\Gamma(1-\eps)}
|M(h)|^2,
\eeq
which can be written as
\beq
s{d\sigma\over dt}(s,t,h)=(1 -h)\ s{d\bar \sigma\over dt}(s,t) +
h\ s{d\sigma_{++}\over dt}(s,t),
\eeq
or as
\beq
s{d\sigma\over dt}(s,t,h)=s{d\bar \sigma\over dt}(s,t) +
h\ s{d\sigma_{LL}\over dt}(s,t),
\eeq
where $sd\bar \sigma/dt$ is the unpolarized cross section and
\beq
s{d\sigma_{LL}\over dt}={1\over2}\left(s{d\sigma_{++}\over 
dt}-s{d\sigma_{+-}\over dt}\right).
\eeq
At the Born level we have,
\beqn
&& s{d\bar \sigma\over dt}(s,t)=e_f^2K_2{\alpha_s\over 
s}T_B(Q^2,u,t), \nonumber \\
&& s{d\sigma_{++}\over dt}(s,t)=-e_f^2K_2{\alpha_s\over 
s}\epsilon A_B(Q^2,u,t),
\eeqn
where
\beqn
&& T_B(Q^2,u,t)={2\over tu}\left[(1-\eps)\left(2Q^2s
+(1-\eps)(t^2+u^2)-2\eps tu\right)\right] \nonumber \\
&& =2(1-\eps)\left[(1-\eps)\left({u\over t}+{t\over u}\right)+
{2Q^2(Q^2-u-t)\over ut}-2\eps\right],
\label{defTB}
\eeqn
and
\beq
A_B(Q^2,u,t)={8\over tu}\left[Q^2s+(t+u)^2\right],
\label{defAB}
\eeq
where $K_2$ is defined by
\beq
K_2=\pi\alpha{C_F\over N_c}
{1\over\Gamma(1-\eps)}
\left({4\pi\mu^2\over Q^2}\right)^\eps\left({sQ^2\over 
tu}\right)^\eps,
\eeq
where we have rescaled, $\alpha_s\to\alpha_s(\mu^2)^\eps$, so 
that it remains dimensionless in $N=4-2\eps$ dimensions.  
At this order of perturbation theory, we have
\beq
s{d\sigma\over dt}(s,t,h)=e_f^2K_2{\alpha_s\over 
s}\left[(1 -h)T_B(Q^2,u,t) -h\epsilon A_B(Q^2,u,t)\right].
\label{bornddt}
\eeq
As we saw earlier, in the limit $\epsilon\to 0$ helicity is
conserved so that
\beq
s{d\sigma\over dt}(s,t,h)=(1 -h)s{d\bar \sigma\over dt}(s,t),
\label{spin2}
\eeq

which implies that 
\beq
s{d\sigma_{++}\over dt}(s,t)=0\ {\rm and}\ s{d\sigma_{LL}\over dt}(s,t)=
-s{d\bar \sigma\over dt}(s,t).
\eeq
To connect with the notation of ref.~\cite{EMP}, we note that
\beq
K_2T_B(Q^2,u,t)=KT_0(Q^2,u,t),
\eeq
where $K$ and $T_0$ are define in ref.~\cite{EMP} as
\beq
K=2\pi\alpha{C_F\over N_c}
{(1-\eps)\over\Gamma(1-\eps)}
\left({4\pi\mu^2\over Q^2}\right)^\eps\left({sQ^2\over 
tu}\right)^\eps,
\eeq
and
\beq
T_0(Q^2,u,t)=\left[(1-\eps)\left({u\over t}+{t\over u}\right)+
{2Q^2(Q^2-u-t)\over ut}-2\eps\right].
\eeq

\section{Virtual diagram contribution}
The list of the diagrams with virtual corrections 
contributing to the non-singlet sector of 
Drell-Yan is given in Fig.~4. We have omitted all the self-energy insertions of quarks and gluons and the ghost contributions. 

Our calculations are performed in the $\overline{MS}$ scheme using dimensional regularization to regulate both the ultraviolet and the infrared 
singularities. The isolation of the (unrenormalized) 
scalar invariant amplitudes at 1 loop level
from the tensor integrals are done using the 
Passarino-Veltman \cite{PV} procedure as implemented in FeynCalc \cite{Mertig}.

We remove the ultraviolet singularities in the relevant subdiagrams by off 
shell regularization. Then by sending on shell the initial state quarks and 
the final state gluon, we encounter singularities in the form of double poles 
and single poles in $\epsilon = 2- n/2 $. 
The one-loop order result for the cross section for the process
$q+\bar{q} \rightarrow \gamma^* + g$
is given by
\begin{eqnarray}
&&s{d\sigma^{\rm virtual}\over dtdu}(s,t,u,h) = \nonumber \\
&&e^2_f K_2\frac{\alpha_s}{s}\delta(s+t+u-Q^2)
\left\{ \left((1-h)T_B - h\epsilon A_B\right) 
\left[ 1 - \frac{\alpha_s}{2\pi}\frac{\Gamma(1-\epsilon)}{\Gamma
(1- 2 \epsilon)}\left(\frac{4\pi \mu^2}{Q^2}\right)^{\epsilon}\right.\right.
\nonumber \\
 &  & \left.\times\left(\frac{2 C_F + N_C}{\epsilon^2} 
+ \frac{1}{\epsilon}
\left(3 C_F - 2 C_F \ln \frac{s}{Q^2} + \frac{11}{6} N_C + N_C \ln
\frac{sQ^2}{ut} -\frac{1}{3}N_F\right)\right)\right]
\nonumber \\
&&+\frac{\alpha_s}{2\pi}(1-h)\left[\pi^2(4C_F + N_C)\frac{ 2Q^2 s + t^2 + u^2}{3tu}
-2(2C_F- N_C)\frac{Q^2 (t^2 + u^2)}{tu(t +  u)}  \right.
\nonumber \\ &&
 -2 C_F  
\left(\frac{8(2Q^2 s +t^2+u^2)}{tu}  -\frac{Q^4 s(t+u)}{t u
(Q^2-u)(Q^2-t)} -\frac{t^2+u^2}{(Q^2-u)(Q^2-t)}\right)   
\nonumber \\ &&
  -2
 \left( Li_2\left(\frac{t}{t-Q^2}\right) +\frac{1}{2}\ln^2\left(1 - \frac{Q^2}{t}\right) 
\right)
\left(N_C \frac{2 s+ t}{u} + 
     2 C_F \frac{s^2+(s+u)^2}{t u}\right) 
\nonumber \\ &&
- 2
  \left(Li_2\left(\frac{u}{u-Q^2}\right) +\frac{1}{2}\ln^2\left(1 - \frac{Q^2}{u}\right)
\right)
\left(N_C \frac{2 s+ u}{t} 
     +2 C_F \frac{s^2+(s+t)^2}{t u}\right) 
\nonumber \\ &&
+ 2 (2 C_F - N_C)\left(
Li_2\left(-\frac{t + u}{s}\right)
 \frac{2Q^2 s +u^2+t^2+2s^2}{t u} \right.
\nonumber \\ &&
+ \left(2\ln\left(\frac{s}{Q^2}\right) 
 \frac{Q^4 - (t+u)^2 }{(t + u)^2} 
 + \ln^2\left(\frac{s}{Q^2}\right)\frac{s^2}{t u} \right)
\nonumber \\ &&
-  \left(\ln\left(\frac{|t|}{Q^2}\right)\ln\left(\frac{s}{Q^2}\right) 
- \frac{1}{2} \ln^2\left(\frac{|t|}{Q^2}\right) \right)
     \frac{s^2+(s+u)^2}{t u} 
\nonumber \\ &&
\left. - \left(\ln\left(\frac{s}{Q^2}\right) \ln\left(\frac{|u|}{Q^2}\right) 
- \frac{1}{2} \ln^2\left(\frac{|u|}{Q^2}\right) \right)
     \frac{s^2+(s+t)^2}{t u} \right)
\nonumber \\ &&
+2 \ln\left(\frac{|u|}{Q^2}\right)\left( C_F\frac{4 Q^2 s -2 s u + t u }
{(Q^2 - u)^2} + N_C \frac{u}{Q^2 - u} \right)
\nonumber \\ &&
+ 2 \ln\left(\frac{|t|}{Q^2}\right) \left(C_F \frac{4Q^2 s-2 s t+t u
 }{(Q^2 - t)^2 } + N_C  \frac{t}{Q^2 - t} \right) 
\nonumber \\ &&
\left.\left.-2\ln\left(\frac{|t|}{Q^2}\right)\ln\left(\frac{|u|}{Q^2}\right) N_C
\frac{2 Q^2 s+t^2+u^2 }{ t u} 
\right]\right\},
\label{mainr}
\end{eqnarray}
where $h=h_1h_2$.
The quantities $T_B$ and $A_B$ appear in at the Born level (\ref{bornddt}) and are given by equations 
(\ref{defTB}) and (\ref{defAB}), respectively. Notice that the double poles and single pole structure (in the $h\to 0$ limit) automatically reproduces the singularities of the virtual contributions of 
ref.~\cite{EMP}.  In particular, the double poles in $\epsilon$ correctly 
multiply the two-to-two Born contribution $T_B$, which is the 
Born level unpolarized result. Furthermore, it is possible to show \cite{Chetal} that the structure of the virtual contributions presented here correctly factorize, after adding the real contributions. 

The presence of the $h\epsilon A_B$ term in (\ref{mainr}) 
implies that in the t'Hooft-Veltman \cite{thv,bm} scheme the virtual 
corrections by themselves do not conserve helicity (\ie\ they do not satisfy the equation (\ref{spin2})).  In particular, the finite part of
$d\sigma_{++}/dt$ in the limit $\epsilon\to 0$ for the virtual corrections is given by 
\beqn
&&s{d\sigma_{++}^{\rm virtual}\over dt}(s,t) = \nonumber \\
&&-e^2_f K_2\frac{\alpha_s^2}{2\pi s}
A_B(s,t)\left(3 C_F - 2 C_F \ln \frac{s}{Q^2} + \frac{11}{6} N_C + N_C \ln
\frac{sQ^2}{ut} -\frac{1}{3}N_F\right),
\label{plusplus}
\eeqn
where $s+t+u=Q^2$.
In some regularization schemes, such as those enforcing an anticommuting $\gamma_5$ in $n$ dimensions, helicity would be conserved and equation (\ref{spin1}) maintained throughout. In the t'Hooft-Veltman \cite{thv,bm} scheme helicity is not manifestly conserved for partial contributions which are not infrared 
safe. 

\section{Real emissions}
Moving to the contributions from the real emissions in the non singlet sector in the same scheme, it is possible to show that they factorize according to the form

\begin{eqnarray}
\!\!&&\!\!\!\!\!\!\!\!\!s\frac{d\sigma^{\rm real}}{dtdu}(s,t,u,h)=
\nonumber \\
&& e^2_f K_2\frac{\alpha_s}{s}
\left\{ \left((1-h)T_B- h\epsilon A_B\right)\delta(s+t+u-Q^2) 
\left[ \frac{\alpha_s}{2\pi}\frac{\Gamma(1-\epsilon)}{\Gamma
(1- 2 \epsilon)}\left(\frac{4\pi \mu^2}{Q^2}\right)^{\epsilon}\right.\right.
\nonumber \\
 &  &\left. \left. \times\left(\frac{2 C_F + N_C}{\epsilon^2} 
+ \frac{1}{\epsilon}
\left(3 C_F - 2 C_F \ln \frac{s}{Q^2} + \frac{11}{6} N_C + N_C \ln
\frac{sQ^2}{ut} -\frac{1}{3}N_F\right)\right)\right]\right\} \nonumber \\  
 & &+(1-h)(d\sigma_1+d\sigma_{II}+d\sigma_{III}) -h\sigma^{hat} -(1 +h)d\sigma_{IV},
\label{ppp}
\end{eqnarray}
where $d\sigma_1$ is the remaining finite part of the real 
$q\bar{q}$ annihilation diagrams denoted by $d\sigma_I$ in ref.~\cite{EMP}, and
$d\sigma_{II}$, $d\sigma_{III}$, and $d\sigma_{IV}$ correspond precisely to the {\it finite} unpolarized cross sections defined in \cite{EMP}.  The term $\sigma^{hat}$ is a regularization dependent term that arises from our use of the t'Hooft Veltman scheme.  A more detailed study will be presented elsewhere \cite{Chetal}, however, the spin dependence of each term is shown. 
It is important to observe that the real term in eq.~(\ref{ppp}) contains the same factor of $(1-h) T_0 - h\epsilon A_B$ as the virtual term in eq.~(\ref{mainr}).  Except for the $\sigma^{hat}$ term helicity is conserved along the quark (or antiquark) lines and the non-singlet order $\alpha_s^2$ 
spin dependent differential cross section can be written as follows: 
\beq
s{d\sigma\over dtdu}(s,t,h)=
(1-h)(d\tilde\sigma_{I}+d\sigma_{II}+d\sigma_{III})
-(1+h)d\sigma_{IV},
\label{spinoff}
\eeq
where $d\tilde\sigma_{I}$, $d\sigma_{II}$, $d\sigma_{III}$, and
$d\sigma_{IV}$ are the {\it unpolarized} cross sections given in ref.~\cite{EMP}. The regularization dependent term $\sigma^{hat}$ can be absorbed into the the spin dependent structure functions leaving a
result that is exactly what is expected from helicity conservation and a complete anticommuting scheme for $\gamma_5$.  However, if this is done then one cannot use the $\overline{MS}$ scheme to calculate the $Q^2$ evolution of the spin dependent structure functions (see ref.~\cite{Chetal}).

\section{Conclusions}

We have presented the parton-level analytical results for 
the next-to-leading order non-singlet virtual plus real corrections to the 
Drell-Yan differential cross-section.
Our unpolarized cross section, obtained from the $h\to 0$ limit of 
(\ref{spinoff}) 
agrees with the previous result of
Ellis, Martinelli and Petronzio \cite{EMP} in the non-singlet sector, which we have recalculated. 
The dependence of the differential cross section on the helicity of the initial state partons is shown explicitly (the spins of the final state partons are summed). Although the calculation is very involved 
due to the presence of chiral projectors in the initial state, the result 
is quite compact and has been presented in a form from which 
cancelation of the infrared, collinear and infrared plus collinear 
singularities is evident. 
Both the polarized initial state and the unpolarized cross sections can be obtained from our result.  A more detailed discussion of our results and of the methods 
developed by us in the analysis of the virtual corrections 
will be presented elsewhere \cite{Chetal}.

A study of the transversity distribution to $O(\alpha_s^2)$ is also forthcoming \cite{CCE}.

\section{Acknowledgments}
S. C. and C.C. thank Alan White and the Theory Group at Argonne for 
hospitality.
C.C. thanks P. Nason and the Theory Division at Cern and G. Marchesini and 
Y. Dokshitzer at the Univ. of Milan for hospitality.
We thank J. Elwood for useful discussions.

\newpage
\section*{Figures}
\begin{center}
\begin{figure}[h]
\epsfxsize=55mm
\centerline{\epsfbox{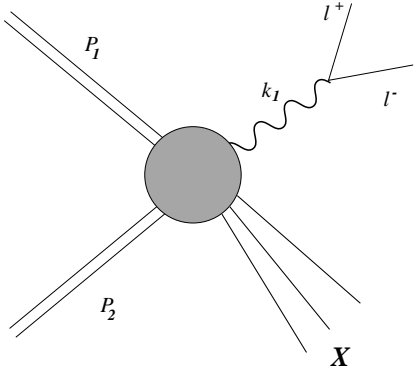}}
\caption{The Drell-Yan process}
\vspace{1cm}
\end{figure}
\begin{figure}[h]
\epsfxsize=140mm
\centerline{\epsfbox{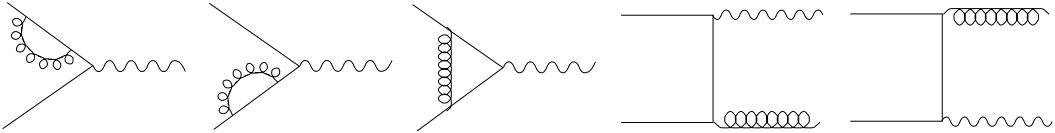}}
\caption{Radiative corrections of order $\alpha_s$}
\end{figure}
\end{center}
\begin{figure}[h]
\epsfxsize=100mm
\centerline{\epsfbox{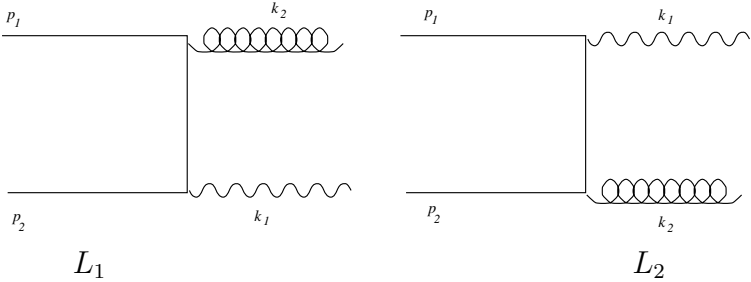}}
\centering{$\!\! L_1$ \hspace{7cm}$\!\! L_2$}
\caption{The Born diagrams}
\end{figure}
\begin{figure}
\epsfxsize=140mm
\centerline{\epsfbox{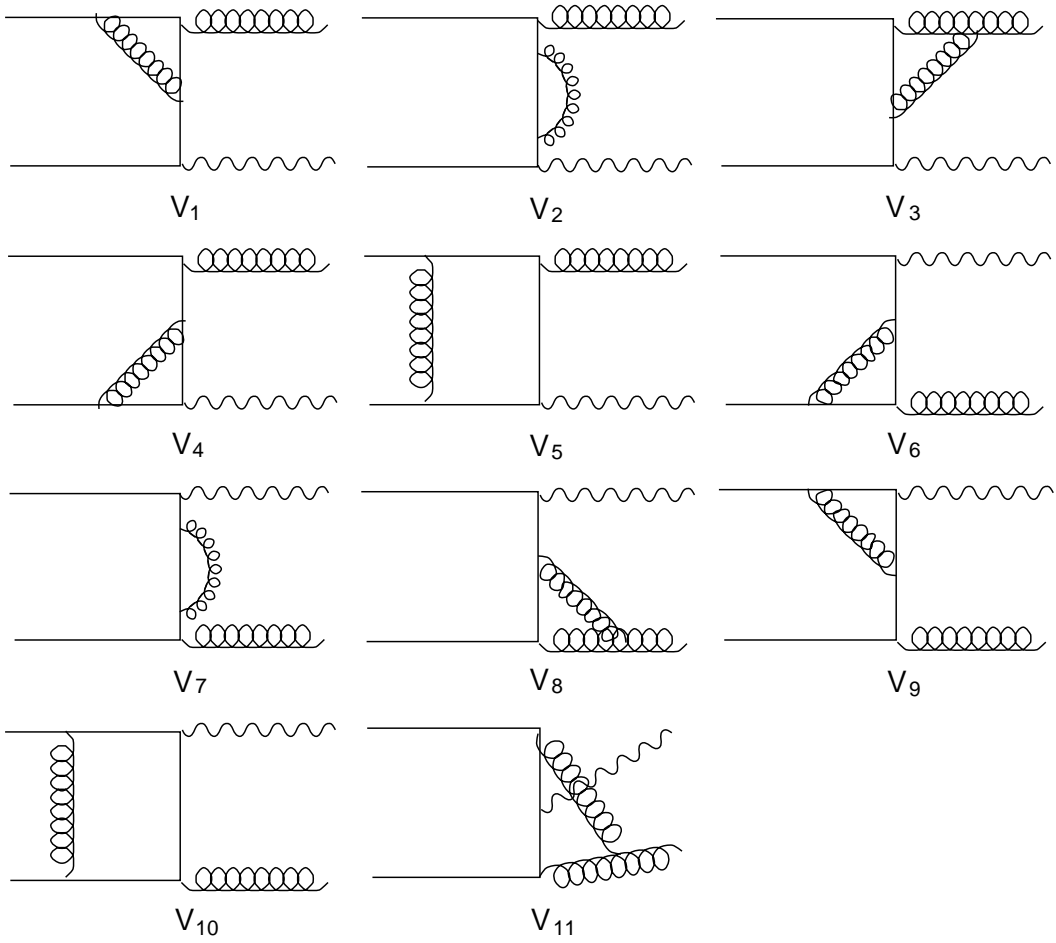}}
\caption{One-Loop contributions}
\end{figure}
\newpage
\begin{figure}
\epsfxsize=70mm
\centerline{\epsfbox{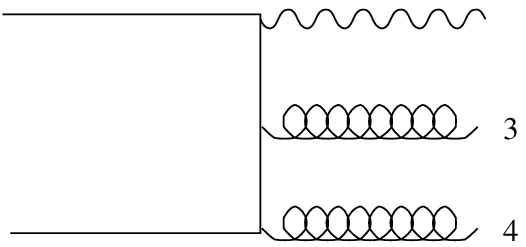}}
\caption{$q+\bar{q} \rightarrow \gamma^* + G +G $}
\end{figure}
\begin{figure}
\epsfxsize=80mm
\centerline{\epsfbox{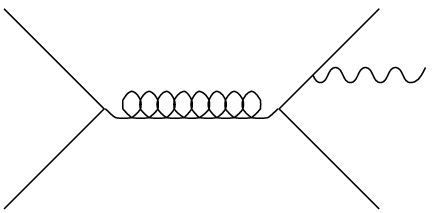}}
\caption{$q+\bar{q} \rightarrow \gamma^* + q+\bar{q} $}
\end{figure}
\begin{figure}
\epsfxsize=70mm
\centerline{\epsfbox{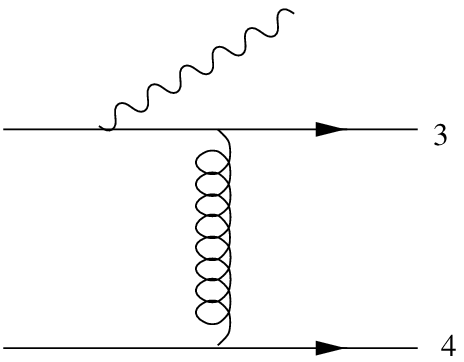}}
\caption{$q+q \rightarrow \gamma^* + q+q $}
\end{figure}

\end{document}